\documentclass[journal]{IEEEtran}
\usepackage{amsmath,amsfonts}
\usepackage{algorithmic}
\usepackage{algorithm}
\usepackage{array}
\usepackage[caption=false,font=normalsize,labelfont=sf,textfont=sf]{subfig}
\usepackage{textcomp}
\usepackage{stfloats}
\usepackage{url}
\usepackage{verbatim}
\usepackage{graphicx}
\usepackage{cite}

\usepackage{microtype}
\usepackage{mathrsfs}
\usepackage{amsthm,amssymb}
\usepackage{multirow}
\usepackage{makecell}
\usepackage{booktabs}

\begin{document}

\title{More Perspectives Mean Better: Underwater Target Recognition and Localization with Multimodal Data via Symbiotic Transformer and Multiview Regression}

\author{Shipei Liu,
Xiaoya Fan,
Guowei Wu}        



\maketitle
\begin{abstract}
Underwater acoustic target recognition (UATR) and localization (UATL) play important roles in marine exploration. 
The highly noisy acoustic signal and time-frequency interference among various sources pose big challenges to this task.
To tackle these issues, we propose a multimodal approach to extract and fuse audio-visual-textual information to recognize and localize underwater targets through the designed Symbiotic Transformer (Symb-Trans) and Multi-View Regression (MVR) method.
The multimodal data were first preprocessed by a custom-designed HetNorm module to normalize the multi-source data in a common feature space.
The Symb-Trans module embeds audiovisual features by co-training the preprocessed multimodal features through parallel branches and a content encoder with cross-attention.
The audiovisual features are then used for underwater target recognition.
Meanwhile, the text embedding combined with the audiovisual features is fed to an MVR module to predict the localization of the underwater targets through multi-view clustering and multiple regression.
Since no off-the-shell multimodal dataset is available for UATR and UATL, we combined multiple public datasets, consisting of acoustic, and/or visual, and/or textural data, to obtain audio-visual-textual triplets for model training and validation.
Experiments show that our model outperforms comparative methods in 91.7\% (11 out of 12 metrics) and 100\% (4 metrics) of the quantitative metrics for the recognition and localization tasks, respectively. 
In a case study, we demonstrate the advantages of multi-view models in establishing sample discriminability through visualization methods.
For UATL, the proposed MVR method produces the relation graphs, which allow predictions based on records of underwater targets with similar conditions.
\end{abstract}


\section{Introduction}
Underwater acoustic target recognition (UATR) and localization (UATL) are important supporting technologies for marine exploration.
Usually, the collected underwater signals are affected by the complex sound field, which forces researchers~\cite{zhou2023novel} to investigate denoising methods to suppress noise.
In addition to denoising, to cope with the characteristics of phase variation with time in underwater acoustic signals, some studies~\cite{sun2021high} are devoted to solving the problem of degradation of recognition accuracy caused by mixing adjacent time slots of received signals with each other.
All these works show that the information loss in unimodal underwater signals will lead to an inevitable decrease in the accuracy of the recognition model.
Several enlightening studies~\cite{SoldiGMBSFTP20} have shown that learning underwater target features from multiple sources can resist environmental interference, which indicates the feasibility of fusing multimodal features of underwater targets.

The primary challenge in fusing multimodal features is eliminating perceptual differences in heterogeneous data.
Considerable efforts have been made by researchers to address this problem.
To integrate cross-modal information of text and images, Yi et al.~\cite{bin2022non} designed parallel non-autoregressive decoders and cross-modal position attention modules in the ordering tasks. 
More multimodal methods are applied to fuse video-audio features, for example, Zhou et al.~\cite{zhou2022mavt} propose a multimodal Transformer to capture the most discriminative regions between audiovisual modalities and outperform other contemporary SOTA methods for image recognition.
Existing underwater data contain visual, acoustic, and implicit knowledge, and we can use experience for reference the work of Xie et al.~\cite{xie2022knowledge}, who proposed a multimodal feature extraction framework based on knowledge augmentation and achieved state-of-the-art (SOTA) performance in video summarization.
Therefore, appropriate methods can effectively fuse cross-modal features, such as synchronizing spatial-temporal information for audiovisual underwater data.

The difference between extracting features from acoustic and visual data is that the spectrum is sampled periodically, while the images are sampled instantaneously.
In some multimodal studies of remote sensing data~\cite{heidler2023self}, prediction models are trained by cropping equal-dimensional features of images and spectrograms, which may lead to learning the visual and acoustic features of the target at different temporal conditions.
These reasons motivate us to model the time series of spectra and images, which can learn the sample variation over time in underwater recognition and localization tasks.
To model consecutive multi-frame images, Lan et al.~\cite{lan2022siamese} proposed a Siamese Transformer network for efficient context propagation from historical frames to current frames in video object segmentation.
Parallel structures, typified by Siamese networks, have also been successfully explored in fusing multimodal features.
The most mature application is fusion using paired multimodal data, as Ozer et al.~\cite{ozer2022siamesefuse} introduced a SiameseFuse model to fuse visible and infrared images.
Many studies~\cite{QuLWS22} have shown that independent task layers with shared parameters can avoid the gradient shrinkage problem, which inspired us to design a branching model with shared parameters for multimodal data.
Therefore, we can leverage the Transformer baseline with the parallel structure to extract the generalized features by estimating the uncertainty of heterogeneous data. 

\begin{figure*}[htbp]
\centering
\includegraphics[width=1.98\columnwidth]{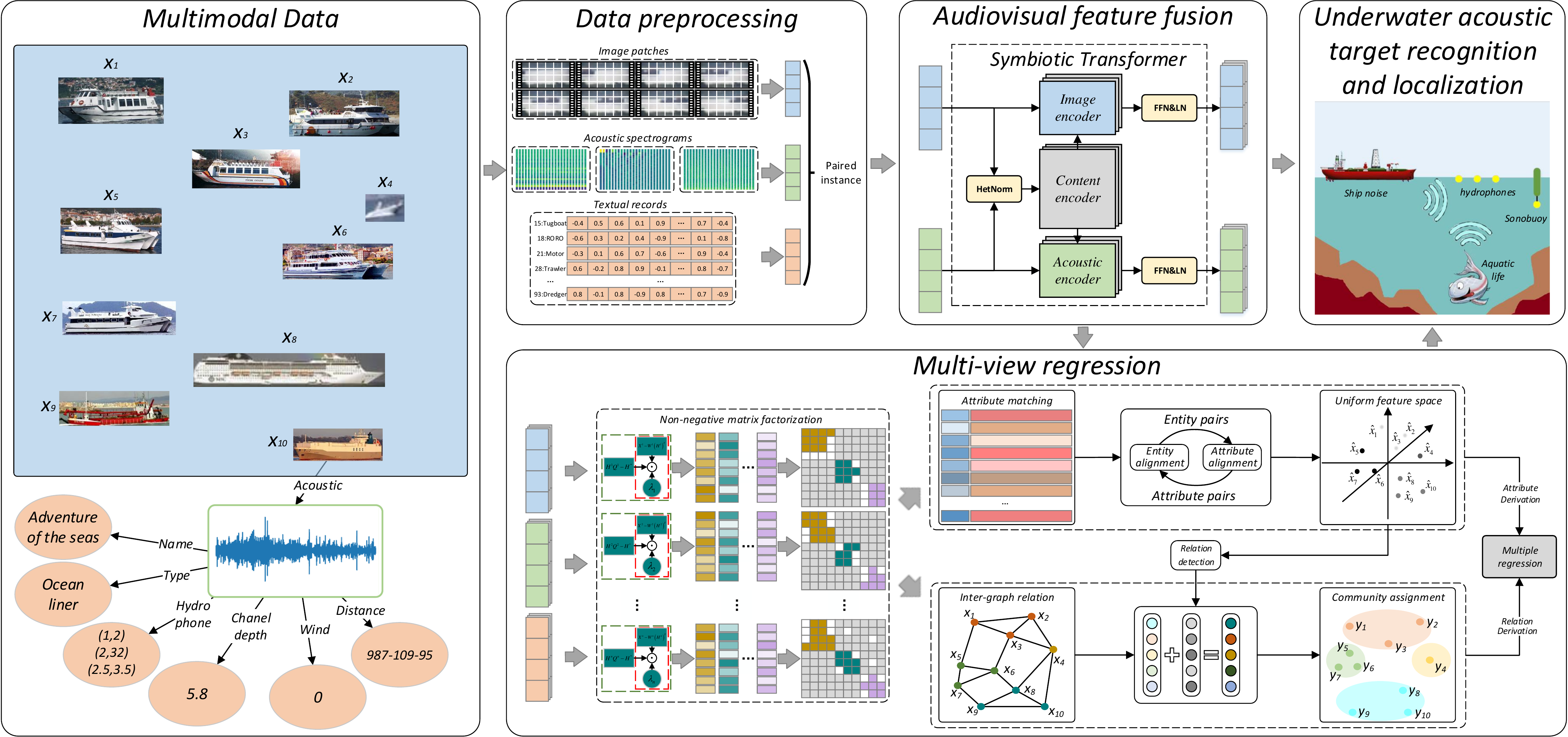}
\caption{Illustration of our multimodal framework for underwater target recognition and localization.
The framework is divided into data preprocessing, heterogeneous feature fusion, and multi-view regression modules.
Video and audio are processed into time-synchronized audiovisual sequences, represented as image patches and spectrograms.
These audiovisual representations are fed into the Symb-Trans model, which performs joint learning through multiple branches for target recognition.
Textual records are processed by an MLP model and combined with audiovisual representations by a multi-view NMF method.
Then, we use multiple regression to predict the distance between targets and sensors based on multi-view clustering and community assignment for target localization.
}
\label{fig1}
\end{figure*}

In this paper, we propose a multimodal method that fuses heterogeneous features to improve the performance of underwater target recognition and localization.
Initially, we preprocess the multi-source data to obtain their features: patch segmentation for video, time-frequency transformation for audio, and word embedding for textual records.
Then, we input audiovisual features into the Heterogeneous Normalization (HetNorm) module to project features to a common space.
These paired representations are fed into a Symbiotic Transformer (Symb-Trans) that optimizes the recognition tasks with a custom-designed loss function.
Finally, in combination with the text embedding of the acquisition environment records, we propose a multi-view regression (MVR) method to locate underwater targets.
An illustration of the proposed framework is shown in Fig. \ref{fig1}. 
Our main contribution can be summarized as follows:

\begin{itemize}
\item We extract multimodal features by combining multiple underwater acoustic and visual datasets, and design a HetNorm module to eliminate the distribution differences between different modalities.
The ablation study showed the effectiveness of the HetNorm module, in particular, the cross-modal normalization performs better than the intra-modal normalization.
\item We propose the Symb-Trans model for recognizing underwater targets, which contains parallel branches for learning audiovisual features.
Experiments show that our model outperforms comparative models in both vision-focused and acoustic-focused underwater recognition tasks.
\item We propose a multi-view regression method based on audio-visual-textual representations to predict the distance between ships and sensors.
Our method references targets with similar conditions detected by community assignment and performs better than comparable models in the localization task.
\end{itemize}

\section{Related works}
\subsection{Underwater acoustic target recognition and localization}
Underwater acoustic target recognition (UATR) and localization (UATL) face the challenge of coping with inaccurate acquisition data, so most methods are dedicated to ameliorating the effects of complex interference.
Under various impaired conditions of the collected data, Doan et al.~\cite{doan2020underwater} proposed a dense CNN model that improves classification accuracy over traditional machine learning models.
In complex underwater acoustic contexts, Wang et al.~\cite{wang2023underwater} proposed an AMNet network with a convolutional attention module to improve the performance of the UATR task.
Transformer-based models are gradually being introduced for underwater target recognition due to their anti-interference and generalization capabilities.
For example, to overcome the bottleneck of capturing global information, Feng et al.~\cite{feng2022transformer} proposed a UATR-Transformer model for perceiving global and local context from the spectrum and improving the accuracy of the UATR task.
Similarly, many excellent deep-learning works have emerged in the UWAL task.
To improve recognition performance by obtaining information from multiple sensors, Meyer~\cite{meyer2017scalable} propose a target tracking algorithm that achieves low computational complexity when computing information from multi-sensors.
For position inference of time-varying targets, Mendrzik~\cite{mendrzik2020joint} proposes a mobile sensors network that uses graphical models to describe the statistical relationship between sensors and targets. 
These works provide the basis and inspiration for learning underwater target features through multimodal and graphical methods.

\subsection{Multimodal feature fusion}
Although deep learning methods have been successfully applied to unimodal classification tasks, they may encounter bottlenecks in fine-grained classification tasks due to the limitation of information diversity.
For example, Audebert et al.~\cite{audebert2019multimodal} explored multimodal models that fuse textual and visual information and learn joint abstract features, achieving better performance than unimodal approaches in document classification and retrieval tasks.
To validate the effectiveness of multimodal frameworks, Hong et al.~\cite{hong2020more} developed a cross-modal architecture and achieved good performance by using fusion features of four remote sensing datasets.
Recently, many researchers have improved the extraction of multimodal features by introducing Transformer baselines.
Most typically, Tsai et al.~\cite{tsai2019multimodal} introduced a multimodal Transformer that uses directional pairwise attention to refer to cross-modal dependencies at different sampling rates, outperforming state-of-the-art unimodal methods on time series prediction tasks.
To train a generic model for handling multimodal tasks, such as video action recognition, audio event classification, and text-to-video retrieval, Akbari et al.~\cite{akbari2021vatt} proposed a video-audio-text Transformer and significantly improved the performance of each task.
Furthermore, to solve the problem of arbitrary matching of the same category in different datasets, Gao et al.~\cite{gao2021adaptive} combined multiple projection tensors for capturing information from homogeneous and heterogeneous data, respectively.

\subsection{Graph-based regression}
Locating the position of vessel targets can be thought of as predicting their distance from monitoring sensors based on multi-view data, such as real-time audiovisual information, and textual records.
A suitable solution for multi-view data prediction is to arbitrarily project the representation of the target onto unlabeled eigenvectors and fit constraints or clustering relationships by intra- or inter-view similarity in the graph model~\cite{wang2018multiview}.
To learn the complex hierarchical relationships in multi-view representations, Huang et al.~\cite{huang2020auto} proposed a deep multi-view clustering model to uncover the hierarchical semantics of the inputting multimodal data in a layer-wise way.
It is effective to fuse multi-view information through partition representations; for example, Zhang et al.~\cite{zhang2021multi} proposed a multi-view clustering algorithm to learn specific view structures and partition-level information and validated the proposed method with six benchmark datasets.
After extracting and fusing multi-view features, the graph-based regression approach provides feasibility for the multidimensional prediction of targets.
A significant study introducing multi-view representations in regression tasks was proposed by Xu et al.~\cite{xu2020learning}, which recasts multi-view depth inference as an inverse depth regression task. 
To measure the positions of the graph elements, we refer to the study by Ferguson et al.~\cite{ferguson2022computation}, which is an approximation algorithm to calculate the Fréchet mean between the eigenvalues of the respective adjacency matrices.

\section{Methodology}
Our approach aims to improve prediction accuracy by combining the complementarity of multimodal features.
We preprocess the multi-source data by image patching, time-frequency conversion, and word embedding to extract heterogeneous features.
By normalizing the audiovisual features with a custom-designed HetNorm module, we propose a Symb-Trans model that optimizes multimodal branching for the underwater target recognition task.
The audiovisual representation is combined with text embedding in our proposed multi-view regression approach to predict the location of underwater targets by community assignment and multiple regression.

\subsection{Data preprocessing}
In the preprocessing stage, we need to convert multi-source data with heterogeneity into feature vectors in a common space, and the difficulty lies in mining the higher-order semantic information.
Therefore, we use different methods to preprocess the multi-source data and convert the results into high-dimensional feature vectors of the same size.

\begin{figure}[htbp]
\centering
\includegraphics[width=0.98\columnwidth]{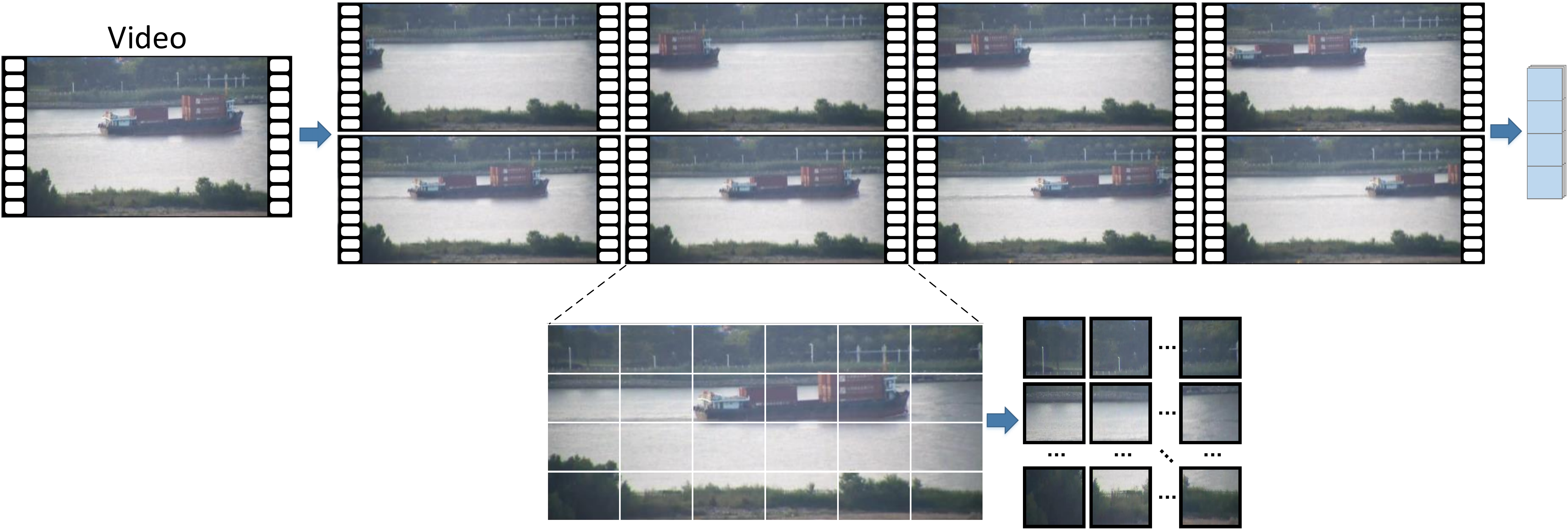}
\caption{Preprocessing image patches from video data.}
\label{fig2}
\end{figure}

\textbf{Image:} 
Benefiting from the visual Transformers (ViT)~\cite{DosovitskiyB0WZ21}, the Transformer baseline is used more often for vision tasks, improving the accuracy of image entity recognition.
ViT models usually divide the image into patches for processing and organize the structural relationships by position embedding (PE).
Considering the different sizes of the images, we use an adaptive pooling method to scale the image to a fixed-size vector.
We first slice the video into time frames, then divide the image by height and width, for example, into 24 (4*6) patches.
The division of video patches is illustrated in Fig. \ref{fig2}.

\begin{figure}[htbp]
\centering
\includegraphics[width=0.98\columnwidth]{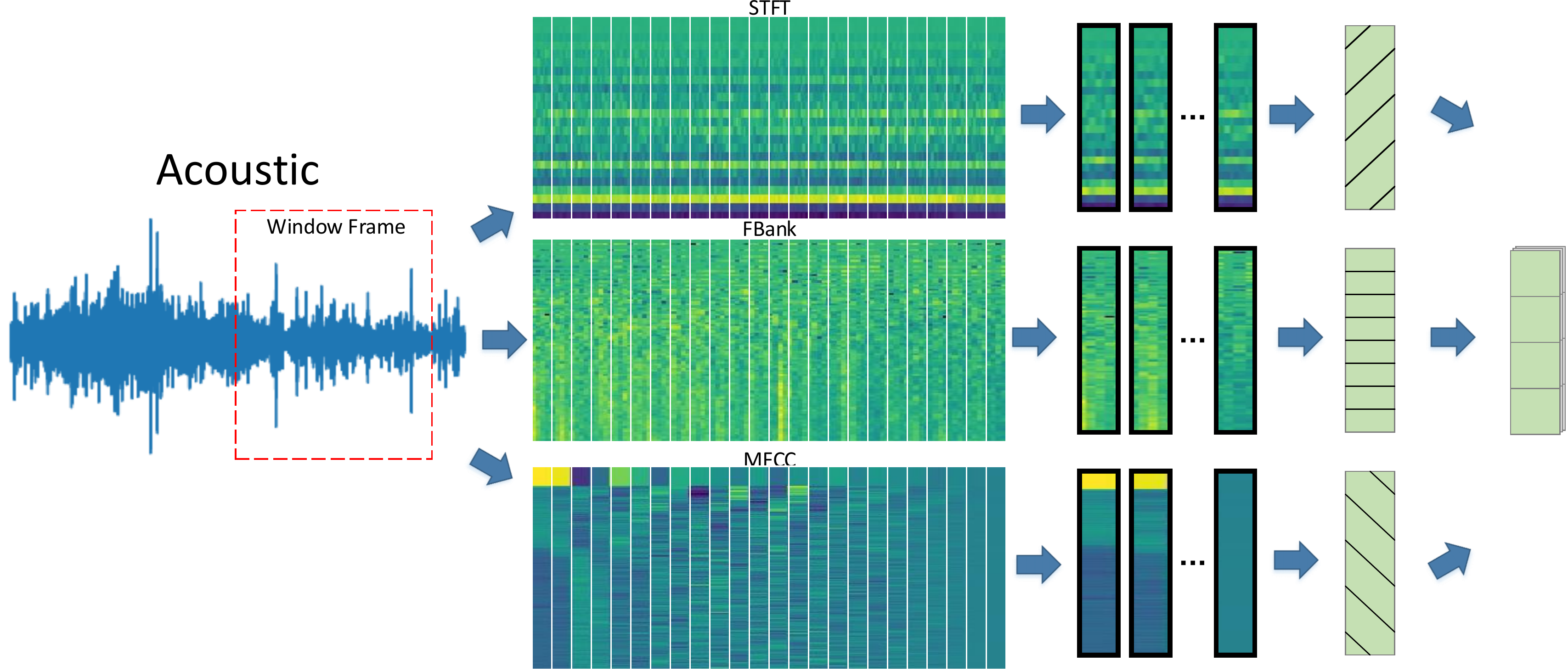}
\caption{Extraction of STFT, FBank, and MFCC spectrograms.}
\label{fig3}
\end{figure}

\textbf{Audio:}
For underwater acoustic signals, short-time Fourier transforms (STFT), Meier filterbank (FBank), and Mel frequency coefficients (MFCC) are commonly used to extract features of audio.
STFT is an extension of the Fourier transform in addressing signal non-stationarity by segmental analysis.
Specifically, the discrete signal is decomposed into smooth segments with a fixed length and then multiplied by a time-limited window function, which is represented as a spectrum by fractional analysis.
In the discrete domain, the STFT of the signal $x(t)$ is defined as:
\begin{equation}
X(w,t)=\sum_{m=-\infty}^{\infty}x(m)g(t-m)e^{-jwt}
\label{eq1}
\end{equation}
where $x(\cdot)$ and $g(\cdot)$ are the input signal and window function, respectively.
FBank uses the Mel scales in the frequency domain, which can represent the strong peak of the signal.
The spectrum is calculated from the Mel filter bank as follows:
\begin{equation}
S_{m}=ln(\sum_{k=0}^{N-1}|X(k)|^{2}\eta_{m}(k)),0\leq m\leq M
\label{eq2}
\end{equation}
where $\eta_{m}(k)$ corresponds to the $m$-th mel filter.
$N$ is the sampling number.
MFCC for the $k$-th frame is calculated by applying discrete cosine transform on the logarithm of magnitude $M_{m}$:
\begin{equation}
C_{n}=\sum_{m=0}^{N-1}S_{m}(M_{m})\cos (\frac{\pi n(m-0.5)}{M}),n=1,\cdots,L
\label{eq3}
\end{equation}
where, $C_{n}$ is the $n$th MFCC spectrum.
In addition, we utilize some experimental tricks, e.g. channel reorganization, to reduce the dimensionality of the spectral time series.
The spectrogram processing results are shown in Fig. \ref{fig3}.

\textbf{Text:}
We use the multilayer perceptron (MLP) model based on bag-of-words (BoW) ~\cite{galke2022bag} to embed textual records of the acoustic dataset.
BoW constructs a vector with dimensions equal to the number of $n_{vocab}$ words in the dictionary.
The BoW-based MLP model, activated by the batch normalization (BN) layer, can be represented by:
\begin{equation}
z=w_{2}\left(BN\left(w_{1}\cdot BoW(x_{text})+b_{1}\right)\right)+b_{2}
\label{eq4}
\end{equation}
where $BoW(x_{text})$ denoted the word bags of the input.
$w$ and $b$ represented the connection factor and the bias, respectively.
$z$ is the embedding of the textual records, and the goal is to learn a generalizable function $z = f_{w,b}(BoW(x_{text}))$ such that $argmax(z)$ is preferably equal to the true value. 

\subsection{Target Recognition: Heterogeneous Feature Fusion}
The prime challenge for heterogeneous feature fusion is the difficulty in modeling with synchronized training in multimodal branches.
We first use the HetNorm module to align heterogeneous samples in the time domain and eliminate their statistical variation.
The details are as follows:

\textbf{Heterogeneous normalization:}
This module was designed to integrate the multimodal training data by stimulating the channel-specific gradients to be of similar magnitude in the feature space. 
The video samples are processed as vectors including dimensions of channel, height, width, and time step, denoted as $X_{video} \in \mathbb R^{CHWT}$.
We construct the corresponding dimensions for spectrogram $X_{audio} \in \mathbb R^{SMBT}$, which are spectral type, amplitude, frequency band, and time step, to match with the video samples.


Assuming that the paired audiovisual instance can represent as $x_{\varphi}\in \mathbb R^{\varphi\sim\phi}$ in a training batch, where $\varphi$ denotes the modal parameter.
The heterogeneity features are normalized to a combined representation by calculating the statistics:
\begin{equation}
HetNorm(x_{\varphi})=\frac{1}{N_{\phi}}\sum_{\varphi}^{\phi}\gamma_{\varphi}\cdot(\frac{x_{\varphi}-\mu(\hat x)}{\sigma(\hat x)})+\beta_{\varphi}
\label{eq5}
\end{equation}
where $N_{\phi}$ is the regularization factor;
$\gamma_{\varphi}$, $\beta_{\varphi}$ are the affine parameters counted in each modal;
$\mu\left(\hat x\right)$ and $\sigma\left(\hat x\right)$ represent the mean and variance of input modal data, respectively.
They are computed independently for mini-batches and channels on each modal to determine the best values during testing:
\begin{equation}
\mu(\hat x)=\frac{1}{N_{\phi}}\sum_{\varphi}^{\phi}x_{\varphi};
\sigma(\hat x)=\sqrt{\frac{1}{N_{\phi}}\sum_{\varphi}^{\phi}(x_{\varphi}-\mu(\hat x))^{2}}
\label{eq6}
\end{equation}
where the mean $\sigma\left(\hat x\right)$ and the variance $\mu\left(\hat x\right)$ can be regarded as scaling factors and translation coefficients.

\begin{figure}[htbp]
\centering
\includegraphics[width=0.98\columnwidth]{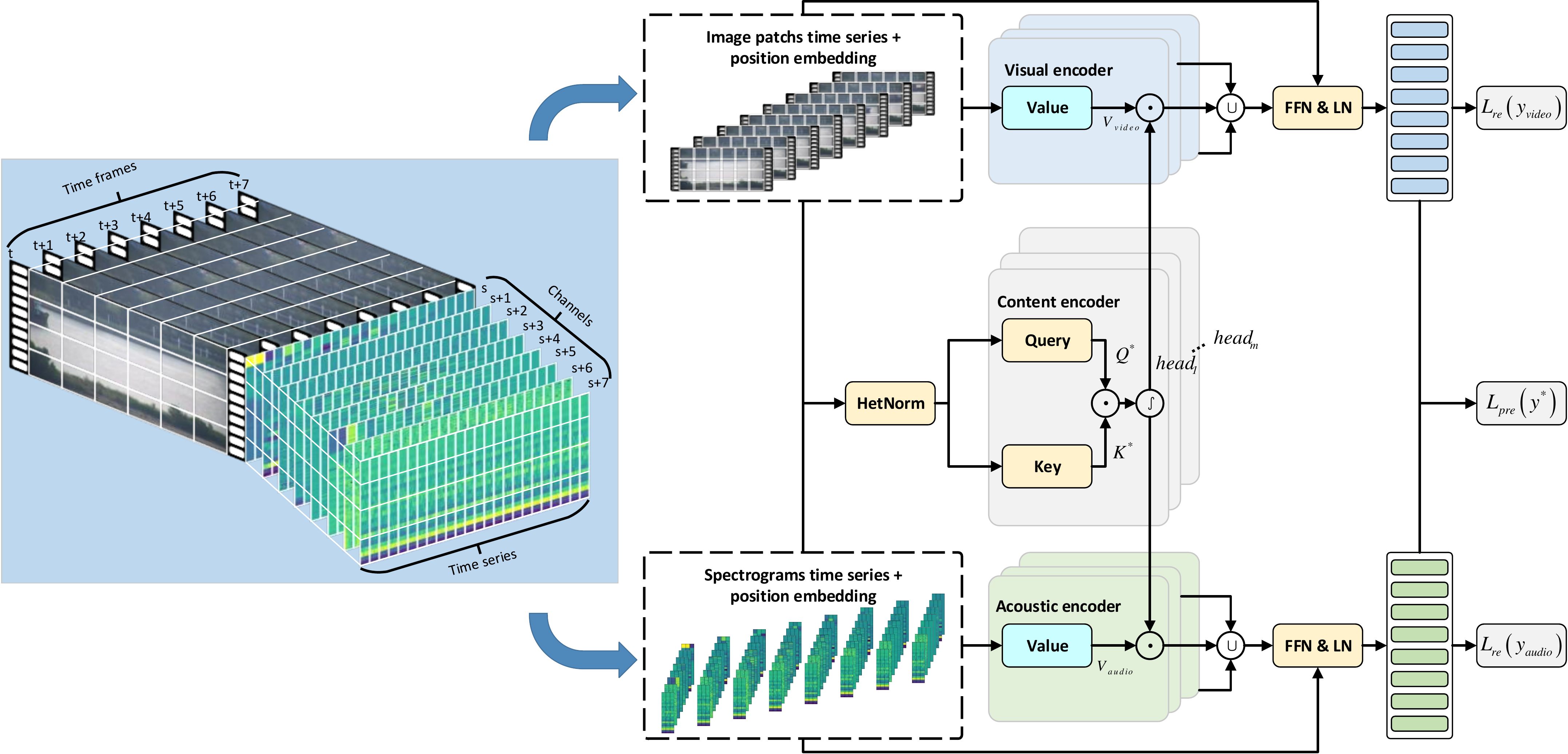}
\caption{Illustration of the proposed Symb-Trans with HetNorm module.
The blue and green blocks represent encoders corresponding to different modalities, and the gray blocks represent contextual content encoders.}
\label{fig5}
\end{figure}

\textbf{Symbiotic Transformer:}
The proposed Symbiotic Transformer model learns heterogeneous features and aggregates them in a parallel structure.
Compared to the Transformer baseline with standard self-attention, it has changed in two ways:
1. The Transformer backbone is divided into a parameter-sharing contextual content encoder and multi-modal branches. 
2. The model is optimized by multiple loss functions, which include two terms, content reconstruction loss, and task-specific prediction loss.
The architecture was detailed in Fig. \ref{fig5}.

We use a multiple channel attention, called MCA, that follows several multimodal branches with canonical self-attention.
The multimodal branches select the focal range through the heterogeneous pattern parameter$\varphi$, denoted as:
\begin{equation}
\begin{aligned}
MCA=\left[\begin{array}{c}
Video: head_{1}\left(x_{\varphi_{v}}\right),\cdots,head_{m}\left(x_{\varphi_{v}}\right)\\
Audio: head_{1}\left(x_{\varphi_{a}}\right),\cdots,head_{m}\left(x_{\varphi_{a}}\right)\end{array}\right]
\end{aligned}
\label{eq7}
\end{equation}
where $x_{\varphi_{v}}$ and $x_{\varphi_{a}}$ represent the visual and acoustic branches of attention;
$m$ is the number of attention headers.
The score of the attention head is a weighted sum of the scaled dot-products of the inputs so that the score of the heterogeneous attention heads can be calculated as:
\begin{equation}
\begin{aligned}
&head(x_{\varphi})=Softmax(\frac{Q^{*} \cdot (K^{*})^{T}}{\sqrt{D}})\cdot V_{\varphi}\\
\end{aligned}
\label{eq8}
\end{equation}
where scaling factor $D$ equals the vector dimension.
$Q^{*}$, $K^{*}$, $V_{\varphi}$ represent the vectors of query, key, and value, respectively.
These feature vectors are transformed by the parameter matrices $W^{Q}$, $W^{K}$, $W^{V}$ through the input $x$, which can calculated by:
\begin{equation}
\begin{aligned}
Q^{*}=&W^{Q}\cdot HetNorm(x_{\varphi});\\
K^{*}=&W^{K}\cdot HetNorm(x_{\varphi});\\
&V_{\varphi}=W^{V}\cdot x_{\varphi}
\end{aligned}
\label{eq9}
\end{equation}
We add the relative position embedding~\cite{WangSLJYLS21} to the multimodal sequences to preserve the temporal information.

The representation $y_{\varphi}$ can be calculated to capture the contextual relation of heterogeneous features, denoted as:
\begin{equation}
y_{\varphi}=LN\left(x_{\varphi}+FFN\left(MCA\left(x_{\varphi}\right)\right)\right)
\label{eq10}
\end{equation}
where $FFN$ and $LN$ represent the Feed-Forward Network and Layer Normalization operation, respectively.
The prediction layer classifies the target based on multimodal representations $y_{\varphi}$, using the Gumbel $Softmax$ function, denoted as $Gumbel(\sum_{\varphi}^{\phi} y_{\varphi}) \to y^{*}$.

\textbf{Multitask loss:} 
Let $X_{pos}(y)= \left[x_{\varphi_{a}}(y), x_{\varphi_{v}}(y^{\prime})\right];y=y^{\prime}$ be a positive instance-label pair and $y$ be the ground-truth label for task-oriented prediction.
To maintain the balance between positive and negative samples, we apply a data expansion trick to select the worst classified samples, denotes as $X_{neg}(y)= \left[x_{\varphi_{a}}(y^{\prime}), x_{\varphi_{v}}(y^{\prime}); y^{\prime}\neq y\right]$. 
We use a contrast loss function to co-train the parameter matrices of the Symb-Trans model with multimodal branches and use a reconstruction loss function to avoid the disappearance of the original information during the optimization process.
Thus, the loss function is designed to two terms, which can be written as:
\begin{equation}
\mathcal{L}_{total}=\lambda\mathcal{L}_{re}(x_{\varphi};y_{\varphi})+\mathcal{L}_{pre}(y^{*})
\label{eq11}
\end{equation}
where $\lambda$ is a parameter used to balance the two loss terms.

The content embedding is sliced as $\mathbb{R}^{\phi\times T}$, where $\phi$ and $T$ represent the number of heterogeneous patterns, and the length of time steps, respectively.
We use the reconstruction loss $\mathcal{L}_{re}$ to construct the correlation between the input $x_{t}$ and output $y_{t}$, which is calculated by a distance-based scoring function:
\begin{equation}
\mathcal{L}_{re}(x_{\varphi};y_{\varphi})=\frac{1}{N_{\phi}}\sum_{\varphi}^{\phi} \left \| \digamma\left(y_{\varphi}\right)-\digamma\left(x_{\varphi}\right) \right \|^{2}_{2}
\label{eq12}
\end{equation}
where the matrix $\digamma$ is a quantitative function used to calculate the correlation between the input and its content embedding.

The prediction layer is optimized by a straight-through estimator that uses Gumbel SoftMax~\cite{JangGP17}. 
By maximizing the likelihood probability based on the multimodal homogeneity, we factorize the output to obtain the prediction loss: 
\begin{equation}
\mathcal{L}_{pre}(y^{*})=-\log \frac{\exp(\left(y^{*}+\theta_{\varphi}\right)/\tau)}{\sum_{\varphi \sim \phi}\exp(\left(y_{\varphi}+\theta_{\varphi}\right)/\tau)}
\label{eq13}
\end{equation}
where $\tau$ is a non-negative spread coefficient;
$y^{*}$ denotes the classification probability;
$\theta$ are samples drawn from $Gumbel(0, 1)$.
The Symb-Trans model outputs two items, a classification probability, and content representations of each modality branch.
Among them, classification probabilities are used to solve the UATR task, and content representations are used to predict the target location.

\subsection{Target localization: multi-view regression}
\begin{figure}[htbp]
\centering
\includegraphics[width=0.8\columnwidth]{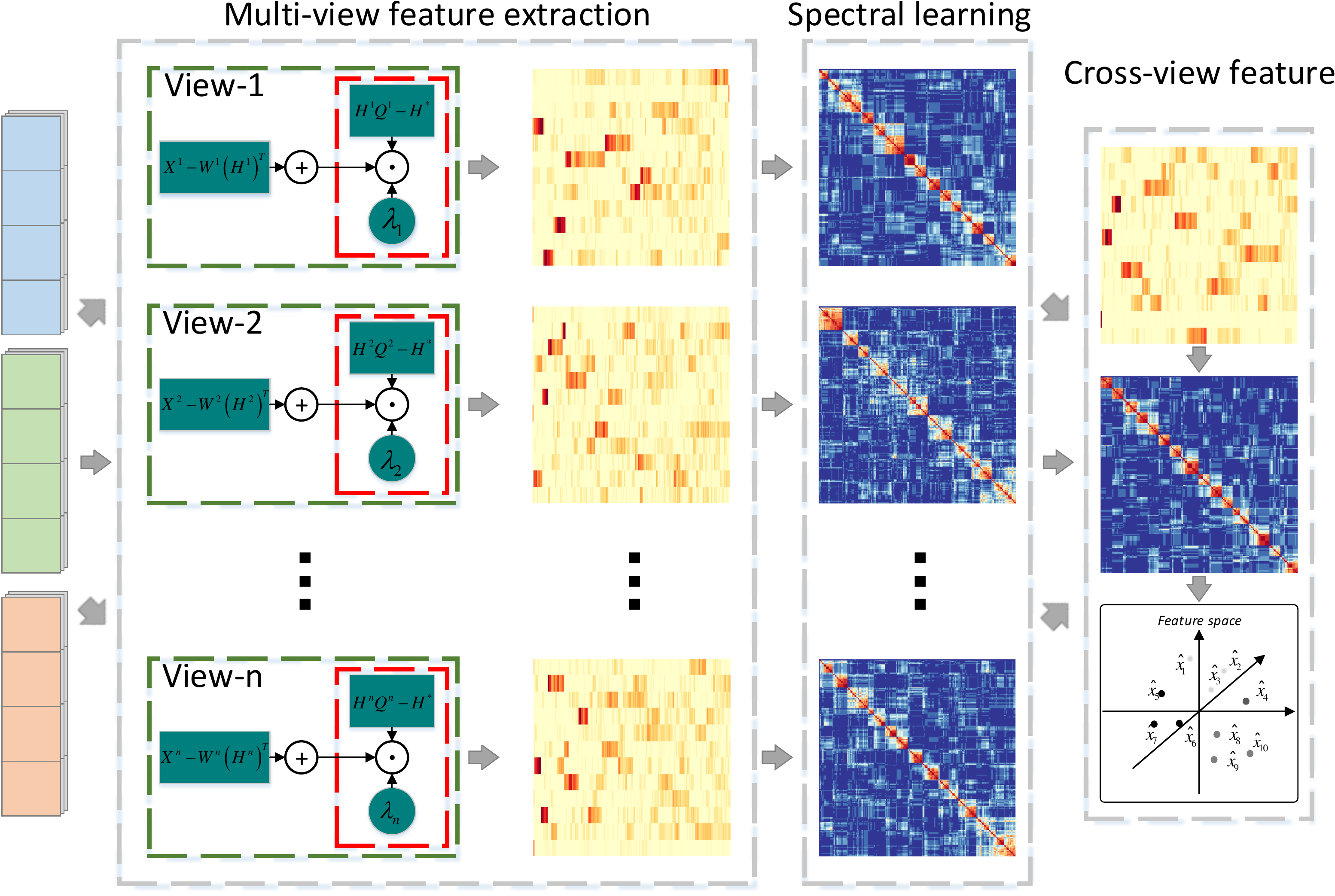}
\caption{Visualization of multi-view NMF processes from audio-visual-textual representations.
Yellow and blue pictures represent the mixture coefficient and consensus matrix.
The reduced-dimensional features are merged into a common space.}
\label{fig6}
\end{figure}

In this section, we propose a multi-view regression (MVR) method to predict the target location.
The MVR method receives the multimodal features learned by the BoW-MLP and Symb-Trans models and predicts the topological relations of the targets.
To refer to the information of higher-order neighbor targets, we construct the relation graph by multi-view clustering and community detection algorithms.
The details of multi-view feature extraction are shown in Fig \ref{fig6}.

\textbf{Multi-view features:}
To perceive the differential features of multimodal data, we handle multi-view representations through a Non-negative matrix factorization (NMF)-based method ~\cite{wang2023generalized}.
The NMF method yields two non-negative low-dimensional matrices whose product approximates the original high-dimensional matrix, intending to map the features of each modality into the same space to advance a common representation compatible with each view.
Given the multi-view representation $X=\{x_{1},\cdots,x_{v}\}\in \mathbb{R}^{M\times S}$ with $v$ views, $S$ and $M$ represent the number of samples and dimensions, respectively.
The objective function is formulated as follows:
\begin{equation}
\begin{aligned}
\min_{W^{v},H^{v},H^{*}}&\sum_{v=1}^{S_{v}}(\big|\big|X^{v}-W^{v}(H^{v})^{T}\big|\big|^{2}_{F}+\lambda_{h}\big|\big|\Delta H^{v}-H^{*}\big|\big|^{2}_{F}),\\
&s.t. \forall 1\leq v\leq S_{v},W^{v}\geq 0,H^{v}\geq 0,H^{*}\geq 0
\end{aligned}
\label{eq14}
\end{equation}
where $||\cdot||_{F}$ represents the Frobenius norm of a matrix;
$\Delta=Diag(\sum_{i=1}^{M}W_{i,1}^{M},\cdots,\sum_{i=1}^{M}W_{i,K}^{M})$ is a diagonal matrix;
$W^{v} \in \mathbb{R}^{M\times K}$ and $H^{v} \in \mathbb{R}^{S\times K}$ are two low-dimensional matrices;
$H^{*} \in \mathbb{R}^{S\times K}$ is the consensus matrix.
The variables $W^{v}$, $H^{v}$, and $H^{*}$ were obtained by iterative updating using the Karush-Kuhn-Tucker (KKT) condition as well as the Lagrange multiplier.
Since $H^{*}$ can be considered as the representation in the low-dimensional space, clustering $H^{*}$ by $argmax_{k}H_{n,k}^{*}$ yields the target relationships.

\textbf{Community Assignment:} 
To provide a more reliable guide to the community assignment of the sample, a graphical model was constructed to quantify the implied relationships.
Given a graph $G = (e,r)$ with the entity set $\{e_{1},\cdots,e_{s}\}$ and the relation set $\{r_{i,j}\}_{s\times s}$, where $i$, $j$ are the iterated variables.
We define a pseudo-metric for computing the distance of neighbor vertices as the $l_{2}$ norm between multi-view representations of neighbor vertex:
\begin{equation}
d_{r}(e_{i},e_{j})=\frac{1}{K}||\psi(H^{*}_{i,k})-\psi(H_{j,k}^{*})||_{2},\forall e\in G
\label{eq15}
\end{equation}
where $H_{i,k}^{*}$ represents the $k$-th view representation of $e_{i}$, and so on.
The function $\psi$ was denoted as a mapping from the set of neighbor matrices $\mathbb{H^{*}}_{s\times s}$ to $\mathbb{R}^{s}$ that assigns to a neighbor matrix the vector of its $s$ sorted eigenvalues.

Communities were determined by minimizing the pseudo-metric between samples $e$ and community $c$, which calculated iteratively using the Frechet mean~\cite{ferguson2022computation}:
\begin{equation}
\mathop{\arg\min}\limits_{e\in c}\frac{1}{N_{c}-1}\sum_{\hat{e} \thicksim c}^{N_{c}-1}d_{r}^{2}(e,\hat{e})
\label{eq16}
\end{equation}
where $N_{c}$ denotes the member number of community $c$.
$\hat{e}$ represents the set of $c$ community members other than $e$, and we need to guarantee that the set has at least one element.

\textbf{Multiple regression:} 
We use a multiple regression method to predict the locations of vessel targets from the multi-view features and the relation matrix.
The multi-view features and the relation matrix are used to train the weights and deviations in the regression equation, in which the dependent variable is determined by multiple independent variables. 
If a target $e$ has $v$ dimension feature and $C$ communities, the predicted values of the target can be regressed as:
\begin{equation}
y(e)=\frac{1}{N_{c}-1}\sum_{\hat{e} \thicksim c}^{N_{c}-1}
\left[\begin{array}{c}
\xi_{1} \\
\ldots \\
\xi_{m}\end{array}\right]\cdot H^{*}(\hat{e})+
\left[\begin{array}{c}
\epsilon_{1} \\
\ldots \\
\epsilon_{m}\end{array}\right]
\label{eq19}
\end{equation}
where $\mathbb{H}^{*}(\hat{e})$ represent multi-view representation in the neighbor matrix with reliable relations selected by vertex $e$ in the $c$-community;
$N_{c}$ represents the number of $c$ communities;
$\xi$, $\epsilon$ represent regression weights and biases, respectively; 
The regressed result $y(e)$ of multi-view representation are a vector, so we superimpose each row into a matrix and introduce a mean squared loss function $\mathcal{L}_{reg}=\frac{1}{M}\sum_{m=1}^{M}(y_{m}-H^{*}_{m})^{2}$ to optimize the regression model.

The proposed MVR method learns multi-view representations and predicts the textual records through a graph-based approach.
In the localization experiments, the effectiveness of our MVR method is verified by predicting the distance between the underwater target and the sampling device.

\section{Experiments and Results}
In naturally collected underwater data, only a few time frames contain information about distinguishable entities, such as engine sounds, and most sound frames are composed of smooth white noise and associated background sounds.
Most importantly, it is necessary to construct a balanced dataset for each category to improve the recall of rare categories. 

\renewcommand\arraystretch{1.2}
\begin{table*}[!htbp]
\centering
\caption{Comparison of our model with SOTA models on classification tasks.
$IntNorm$ denotes intra-modal normalization.
ACC$_{1s}$, UAR$_{1s}$, F-1$_{1s}$ represent metrics calculated in spectrums segmented by 1 second, respectively. 
The chance level is 50.0\% UAR for the classification tasks. 
The best result for each metric is highlighted in bold.}
\label{table1}
\resizebox{1.98\columnwidth}{!}{
\begin{tabular}{ccccccccccccccc}
\hline
\multirow{3}{*}{\textbf{Method}}&\multirow{3}{*}{\textbf{\makecell[c]{Training}}}&\multirow{3}{*}{\textbf{\makecell[c]{Model\\config}}}&\multicolumn{6}{c}{\textbf{\makecell[c]{DeepShip-Seaship(4 categories)}}}&\multicolumn{6}{c}{\textbf{\makecell[c]{ShipsEar-SeaShip(6 categories)}}}\\
&&&\multicolumn{3}{c}{\textbf{\makecell[c]{Acoustic classification}}}&\multicolumn{3}{c}{\textbf{\makecell[c]{Visual classification}}}&\multicolumn{3}{c}{\textbf{\makecell[c]{Acoustic classification}}}&\multicolumn{3}{c}{\textbf{\makecell[c]{Visual classification}}}\\
\cmidrule(r){4-6}\cmidrule(r){7-9}\cmidrule(r){10-12}\cmidrule(r){13-15}
~&~&~&\textbf{ACC$_{1s}$}&\textbf{UAR$_{1s}$}&\textbf{F-1$_{1s}$}&\textbf{ACC}&\textbf{UAR}&\textbf{F-1}&\textbf{ACC$_{1s}$}&\textbf{UAR$_{1s}$}&\textbf{F-1$_{1s}$}&\textbf{ACC}&\textbf{UAR}&\textbf{F-1}\\
\hline
\multirow{1}{*}{AMNet}&\multirow{3}{*}{\makecell[c]{Spectrums}}&\multirow{3}{*}{\makecell[c]{-}}&93.60&90.02&91.78&-&-&-&90.80&87.60&89.17&-&-&-\\
\multirow{1}{*}{STM}&&&96.74&94.92&95.82&-&-&-&92.70&90.44&91.56&-&-&-\\
\multirow{1}{*}{UATR-Trans}&&&97.21&95.57&96.38&-&-&-&94.57&92.18&93.07&-&-&-\\
\hline
\multirow{1}{*}{ImageNet}&\multirow{3}{*}{\makecell[c]{Images}}&\multirow{3}{*}{\makecell[c]{-}}&-&-&-&90.35&87.89&89.1&-&-&-&90.60 &89.07 &89.83\\
\multirow{1}{*}{Xception}&&&-&-&-&91.82&88.90&90.33&-&-&-&93.33 &89.38 &91.31 \\
\multirow{1}{*}{ViT}&&&-&-&-&\textbf{95.82}&93.33&94.56&-&-&-&94.60 &93.32 &93.95 \\
\hline
\multirow{1}{*}{VGG-16}&\multirow{3}{*}{\makecell[c]{Unimodal images\\and spectrums}}&\multirow{3}{*}{\makecell[c]{-}}&92.60&86.62&89.51&89.57&89.52&89.54&88.10&83.20&85.58&89.29 &90.06 &89.67 \\
\multirow{1}{*}{ResNet-50}&&&93.34&90.88&92.09&90.88&89.67&90.27&91.63&88.89&90.24&91.48 &89.88 &90.67\\
\multirow{1}{*}{Transformer}&&&94.75&93.18&93.96&93.61&92.82&93.21&93.53&90.54&92.01&92.64 &91.05 &91.84\\
\hline
\multirow{3}{*}{\makecell[c]{Symb-Trans}}&\multirow{3}{*}{\makecell[c]{Multimodal \\paired instances}}&-&94.75&93.18&93.96&90.85&89.27&90.05&89.27&87.71&88.48&92.10 &87.77 &89.88 \\
~&~&IntNorm&97.22&93.78&95.47&94.69&90.81&92.71&93.59&90.85&92.20&92.88 &90.34 &91.60 \\
~&~&HetNorm&\textbf{97.55}&\textbf{97.36}&\textbf{97.45}&94.78&\textbf{94.92}&\textbf{94.84}&\textbf{96.90}&\textbf{93.98}&\textbf{95.42}&\textbf{94.97}&\textbf{94.16}&\textbf{94.56}\\
\hline
\end{tabular}}
\end{table*}

\subsection{Dataset}
We use acoustic and visual underwater datasets paired for model training and validation, including two acoustic datasets and one visual dataset.
The training and testing sets were divided according to the ratio of 8:2.
Here is a detailed description of each data:

\subsubsection{Deepship~\cite{irfan2021deepship}}
This acoustic dataset consists of more than 47 hours of real-world underwater records of 265 different ships belonging to \textbf{4} categories. 
It provides a record of the distance between the ship and the sensor at 3 different times to better understand the trajectory the ship is following.

\subsubsection{ShipsEar~\cite{santos2016shipsear}}
This acoustic dataset includes 90 long-term audio records with \textbf{12} categories and provides information in structured text, image, and video formats.
The ShipsEar database contains detailed attributes for obtaining each record, e.g., vessel type, location, weather conditions, etc.

\subsubsection{Seaship~\cite{shao2018seaships}}
This dataset designed to evaluate ship target detection algorithms consists of 31,455 images from 10,080 real-world videos covering \textbf{6} ship categories.

\textbf{Audiovisual training materials:} 
To train the multimodal branch of the Symb-Trans model, we construct paired audiovisual instances by excerpting audio and video samples.
Specifically, we combine samples of the same category in the audiovisual dataset into a single paired instance by aligning the sampling window of the spectrogram with the video frames in the same time frame.
This method merges the audio dataset and video dataset one by one, denoted as Deepship-Seaship and ShipsEar-Seaship, respectively.
Details of the audiovisual training materials are described in Appendix B.

\subsection{Underwater target recognition and localization}
We compare the proposed Symb-Trans and MVR models with other models widely used for underwater target identification and localization.
In all experiments, we reported the mean by performing 5 calculations on different random seeds.
Details of the model training are described in Appendix C.

\textbf{Recognition (Acoustic vs Visual classification):} 
Our Symb-Trans and comparative methods are validated in acoustic and visual classification tasks.
We retrained acoustic recognition models, such as AMNet~\cite{wang2023underwater}, STM~\cite{li2022stm} and UATR-Transformer~\cite{feng2022transformer}, on acoustic modality. 
As well as retrained visual detection models, such as ImageNet~\cite{deng2009imagenet}, Xception~\cite{chollet2017xception}, and ViT~\cite{DosovitskiyB0WZ21}, on visual modality.
Considering a generalizable model, we retrained some baselines in two modalities, i.e., VGG~\cite{SimonyanZ14a}, ResNet~\cite{he2016deep} and Transformer~\cite{vaswani2017attention}.
Our Symb-Trans model was trained with Transformer encoders that followed the ViT model configuration.
The ablations were performed to analyze the improvements of the designed HetNorm layer.
We used three quantitative metrics, such as accuracy (ACC), unweighted average recall (UAR), and F-1 score, to analyze the results, as shown in Tab. \ref{table1}.

In general, unimodal models perform better than generic models with the same parameters, which can be attributed to the faster convergence of the unimodal distribution.
For example, the f-1 score of UATR-Trans was better than Transformer on acoustic classification (96.38\% vs. 93.96\% on DeepShip-SeaShip and 93.07\% vs. 92.01\% on ShipsEar-SeaShip).
The same is true compared with ViT on visual classification (94.56\% vs. 93.21\% on DeepShip-SeaShip and 93.95\% vs. 91.84\% on ShipsEar-SeaShip).
Our Symb-Trans model performs better (F-1 scores of 97.45\%, 94.84\%, 95.42\%, and 94.56\% for each task, respectively) than other compared generalization models.
Meanwhile, ablation studies demonstrate the effectiveness of our HetNorm module, especially the cross-modal normalization.
These phenomena can be explained by eliminating statistical differences in multimodal data, which leads to learning nondiscriminatory features in parallel branches.

\textbf{Localization:} 
We evaluated the proposed MVR method in UWTL tasks that were performed with distance and other textual records from two acoustic datasets.
The contrasting SVR, ResNet, and Transformer baselines were retrained on each dataset, and we refer to these initialized models as UWTL-SVR, UWTL-Res, and UWTL-Trans, respectively.
In addition, We test the embedding performance of the WideMLP baseline at different word bag sizes by adding a linear regression layer.
Two common regression metrics, including root mean squared error (RMSE) and coefficient of determination ($R^{2}$), were used to measure the experimental results.

\renewcommand\arraystretch{1.2}
\begin{table}[!htbp]
\centering
\caption{Comparison of models with regression tasks on DeepShip and ShipsEar datasets.
$Bs$ stands for word bag size.
The best result for each task is highlighted in bold.}
\label{table2}
\resizebox{0.98\columnwidth}{!}{
\begin{tabular}{cccccccc}
\hline
\multirow{2}{*}{\textbf{Method}}&\multirow{2}{*}{\textbf{\makecell[c]{Input}}}&\multicolumn{2}{c}{\textbf{\makecell[c]{Config}}}&\multicolumn{2}{c}{\textbf{\makecell[c]{DeepShip-Seaship}}}&\multicolumn{2}{c}{\textbf{\makecell[c]{ShipsEar-Seaship}}}\\
\cmidrule(r){3-4}\cmidrule(r){5-6}\cmidrule(r){7-8}
~&~&Baseline&Bs&RMSE&$R^{2}$&RMSE&$R^{2}$\\
\hline
\multirow{1}{*}{UWTL-SVR}&\multirow{3}{*}{\makecell[c]{Structural\\text}}&-&-&15.81&0.505&4.22&0.436\\
\multirow{1}{*}{UWTL-Res}&~&-&-&13.85&0.629&3.20&0.505\\
\multirow{1}{*}{UWTL-Trans}&~&-&-&13.28&0.643&2.73&0.551\\
\hline
\multirow{3}{*}{WideMLP}&\multirow{3}{*}{\makecell[c]{BoWs}}&\multirow{3}{*}{\makecell[c]{Linear\\regression\\layer}}&1000&22.50&0.337&6.11&0.514\\
~&~&~&2000&19.75 &0.381&5.82&0.523\\
~&~&~&4000&16.96 &0.494&5.26&0.544\\
\hline
\multirow{6}{*}{\makecell[c]{MVR}}&\multirow{3}{*}{\makecell[c]{BoWs+\\single-view\\features}}&\multirow{3}{*}{\makecell[c]{WideMLP+\\Trans}}&1000&18.71 &0.333&4.76&0.464\\
~&~&~&2000&13.76 &0.507&4.24&0.543\\
~&~&~&4000&7.57 &0.667&3.66&0.549\\
\cmidrule{2-8}
~&\multirow{3}{*}{\makecell[c]{BoWs+\\multi-view\\features}}&\multirow{3}{*}{\makecell[c]{WideMLP+\\Symb-\\Trans}}&1000&16.89&0.334&3.87&0.550\\
~&~&~&2000&12.39 &0.656&3.22&0.558\\
~&~&~&4000&\textbf{5.58}&\textbf{0.764}&\textbf{2.29}&\textbf{0.601}\\
\hline
\end{tabular}}
\end{table}

The regression results are recorded in Tab. \ref{table2}, where the proposed MVR outperforms all competitors on both two acoustic datasets.
Specifically, the three models using the WideMLP baseline exhibit a gradual improvement in performance as the word bag size increases.
Compared to the single-view feature, the MVR model using the multi-view feature exhibited smaller regression errors for the same configuration, while far outperforming the other comparative models.
Among them, the MVR model reaches the best value of all models at $cs=4000$.

\subsection{Case studies}
In addition to quantitative analysis of recognition and localization tasks, we also investigate the detailed processes of multi-view representation, community assignment, and relationship graph generation on the ShipsEar dataset.

\begin{figure}[htbp]
\centering
\includegraphics[width=0.98\columnwidth]{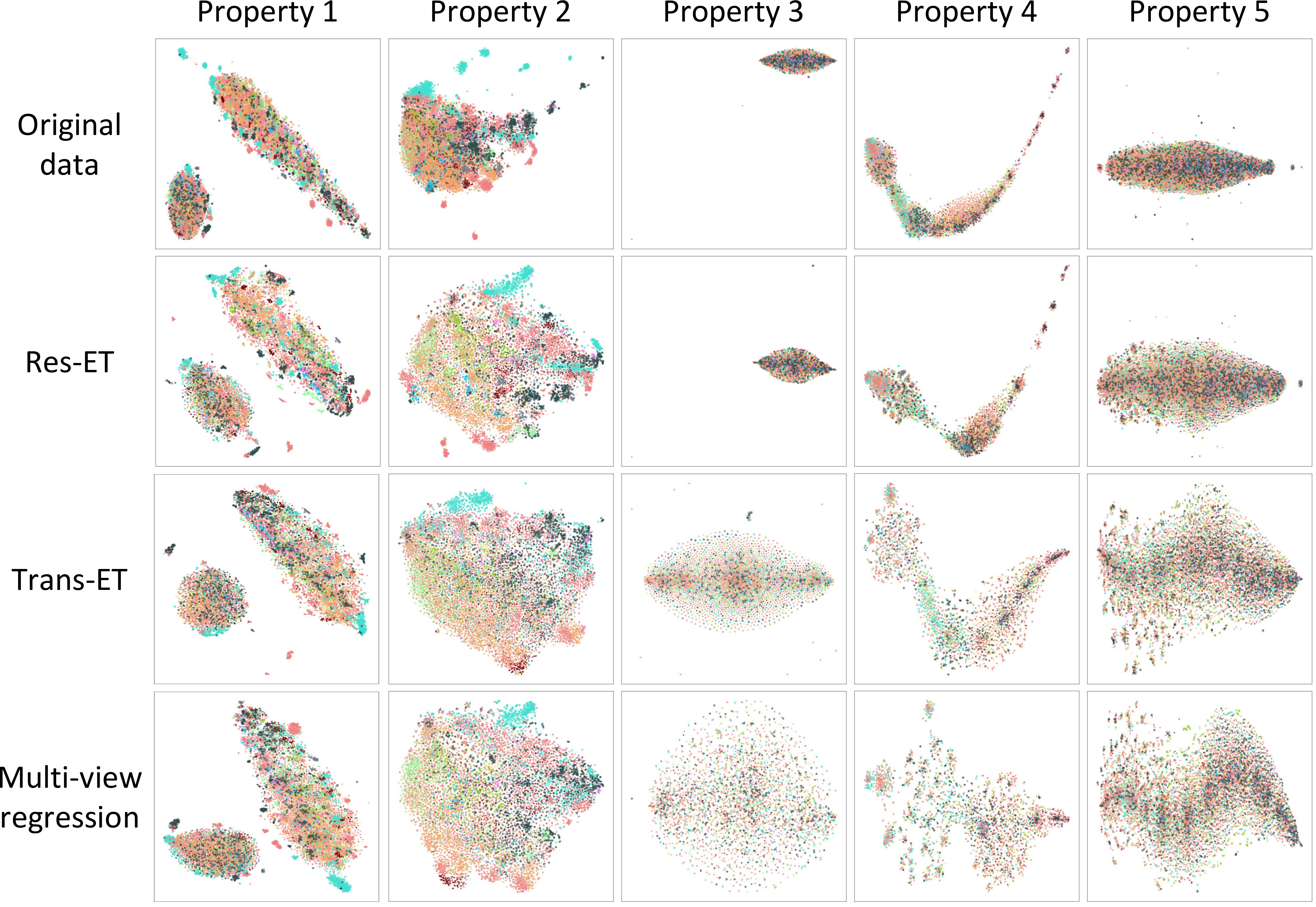}
\caption{Feature representations visualized with t-SNE method on the ShipsEar dataset.
Two-dimensional visualization on five properties.}
\label{fig8}
\end{figure}

\textbf{Feature visualization:}
We visualize the representations of different dimensions by employing the t-SNE algorithm to visualize the superiority of the MVR method in generalizing the differentiability of features.
The distance between adjacent nodes is getting more substantial, indicating that models can reveal fine-grained node relationships and ground-truth flat partitions from representation learning.
Each row represents the distribution of original data, the UWTL-Res representation, the UWTL-Trans representation, and our MVR representation, respectively.
As shown in the Fig. \ref{fig8}, each column represents a property of textual records, obtained from the embedding baselines.

As we can see, the feature representations generated by UWTL-Res cannot be separated effectively, especially the third and fifth attributes, which can lead to difficulties in the successor clustering task.
Compared with the single-view approach UWTL-Trans, our model further enhances the scalability of information and thus improves the distinguishability of nodes.
In particular, our MVR approach reduces the overlap of different categories and allows nodes within the same category to be close to each other.

\textbf{Community assignment:}
We identify target relationships in multi-view space by feature mapping and clustering and validate their performance by calculating the difference between the predicted community and the ground truth.
We train UWTL-Res and UWTL-Trans after removing the corresponding network layers and then use these models without multi-view branches as a comparison.
The clustering results are evaluated by two metrics, such as the graph limits~\cite{lovasz2012large}, which represent distinguishability, and the graph kernels~\cite{nikolentzos2021graph}, which represent similarity.
As shown in Tab. \ref{table3}, the proposed multi-view modules contribute to the community assignment performance in almost all categories on the ShipsEar dataset.

\renewcommand\arraystretch{1.2}
\begin{table}[htbp]
\centering
\caption{Comparison of community assignment results from three models on the ShipsEar dataset.
$GLimit$ and $GKernel$ represent the graph limits and graph kernels of the detected communities relative to the original communities.
The best result for each community is highlighted in bold.}
\label{table3}
\resizebox{0.98\columnwidth}{!}{
\begin{tabular}{cccccccc}
\hline
\multirow{2}{*}{\textbf{Community}}&\multirow{2}{*}{\textbf{Number}}&\multicolumn{2}{c}{\textbf{\makecell[c]{UWTL-Res}}}&\multicolumn{2}{c}{\textbf{\makecell[c]{UWTL-Trans}}}&\multicolumn{2}{c}{\textbf{\makecell[c]{MVR}}}\\
\cmidrule(r){3-4}\cmidrule(r){5-6}\cmidrule(r){7-8}
~&~&GLimit&GKernel&GLimit&GKernel&GLimit&GKernel\\
\hline
Dredger&262 (5)&0.854 &0.455 &0.967 &0.556 &\textbf{0.989}&\textbf{0.989}\\
Passengers&3981 (30)&0.549 &0.482 &\textbf{0.992}&0.394 &0.975 &\textbf{0.670}\\
Tugboat&200 (2)&\textbf{1.000}&0.624 &0.952 &0.621 &\textbf{1.000}&\textbf{0.987}\\
Ocean liner&842 (7)&0.965 &0.402 &0.977 &0.141 &\textbf{0.996}&\textbf{0.445}\\
RORO&1500 (5)&0.981 &0.286 &0.970 &\textbf{0.456}&\textbf{0.998}&0.315 \\
MotorBoat&1007 (13)&0.786 &0.273 &\textbf{0.985}&0.634 &0.972 &\textbf{0.664}\\
Trawler&163 (1)&\textbf{1.000}&0.926 &0.936 &0.926 &\textbf{1.000}&\textbf{0.964}\\
Pilot ship&136 (2)&\textbf{1.000}&0.353 &0.954 &\textbf{0.586}&\textbf{1.000}&\textbf{0.586}\\
Sailboat&409 (4)&0.955 &0.325 &\textbf{0.962}&0.331 &0.995 &\textbf{0.333}\\
Mussel boat&662 (5)&0.830 &0.125 &0.967 &\textbf{0.394}&\textbf{0.983}&0.113 \\
Fish boat&516 (3)&0.855 &\textbf{0.535}&0.959 &0.426 &\textbf{0.981}&0.263 \\
Natural noise&1092 (12)&0.810 &0.301 &0.980 &0.329 &\textbf{0.987}&\textbf{0.857}\\
\hline
Overall&10773 (90)&0.882 &0.424 &0.967 &0.483 &\textbf{0.990}&\textbf{0.599}\\
\hline
\end{tabular}}
\end{table}

\textbf{Relation graph:}
We constructed relationship graphs from community assignment results on the ShipsEar dataset.
Comparison experiments were performed on the original data, UWTL-Trans, and our MVR method to extract relationships separately.
Eight quantitative metrics were used to evaluate the constructed relationship graphs, as detailed in Appendix D.

\renewcommand\arraystretch{1.2}
\begin{table}[htbp]
\centering
\caption{Comparison of the evaluation indicators of knowledge graphs generated from the dataset, UWTL-Trans, and our MVR model.
The best result for each task is highlighted in bold.
}
\label{table4}
\resizebox{0.98\columnwidth}{!}{
\begin{tabular}{cccc}
\hline
\multicolumn{1}{c}{\textbf{Metrics}}&\textbf{ShipsEar}&\textbf{UWTL-Trans}&\textbf{\makecell[c]{MVR}}\\
\hline
Average degree&4.51&5.20$\uparrow$ \scriptsize{(0.69)}&4.55$\uparrow$ \scriptsize{(\textbf{0.04})}\\
Diameter&6&5&6\\
Clustering coefficient&2.1$e$-03&1.64$e$-02$\uparrow$ \scriptsize{(1.43$e$-02)}&5.4$e$-03$\uparrow$ \scriptsize{(\textbf{3.3$e$-03})}\\
Graph density&1.28$e$-02&1.51$e$-02$\uparrow$ \scriptsize{(2.3$e$-03)}&1.303$e$-02$\uparrow$ \scriptsize{(\textbf{2.3$e$-04})}\\
Closeness centrality&0.289&0.297$\uparrow$ \scriptsize{(8$e$-03)}&0.29$\uparrow$ \scriptsize{(\textbf{1$e$-03})}\\
Betweenness centrality&7.2$e$-03&7.1$e$-03$\downarrow$\scriptsize{(1$e$-04)}&7.2$e$-03(\textbf{-})\\
Transitivity&6.3$e$-03&3.26$e$-02$\uparrow$\scriptsize{(2.63$e$-02)}&1.503$e$-02$\uparrow$\scriptsize{(\textbf{8.73$e$-03})}\\
Effective size&3.531&3.443$\downarrow$\scriptsize{(8.8$e$-02)}&3.516$\downarrow$\scriptsize{(\textbf{1.5$e$-02})}\\
\hline
\end{tabular}}
\end{table}

As shown in Tab. \ref{table4}, a higher degree indicates that the connectivity of the nodes in the graph has deteriorated, i.e., there is an implied relationship that has not been identified.
Shorter diameters mean that fewer graphical elements are perceived, which causes a chain reaction in the UWTL-Trans results: higher clustering coefficients, higher graphical density, and higher closeness centrality.
Higher betweenness centrality indicates a higher number and size of public nodes, which predicts that the clusters in our graph are more obviously placed in the same area.
Transitivity equal to the global clustering coefficient is more focused on height nodes. 
The efficiency of the UWTL-Trans graph (calculated by dividing the effective size by the node degree) is rapidly reduced compared to our method.
In summary, compared to the UWTL-Trans model, our MVR approach constructs a relational graph that is closer to the original dataset, which can be interpreted as avoiding relational omissions of detected targets.

\section{Discussion and Conclusion}
Due to the immaturity of multimodal data acquisition techniques, most UATR and UWTL methods utilize unimodal sampling rather than multimodal.
We combine multiple datasets to combine multimodal time series features for identification and achieve better performance than unimodal models.
To solve the problem of multimodal features fusion, we propose: 

1) a normalization method for eliminating differences in the distribution of heterogeneous features;

2) a Transformer model with parallel branches, called Symb-Trans, embedding audiovisual samples into high-dimensional matrices for underwater target recognition;

3) a multiple regression method that can fuse audiovisual features with textual information to jointly infer the distance between the underwater target and the hydrophone by describing adjacent vertices through a multi-view representation.

Experimental results show that the Symb-Trans model outperforms the SOTA models on the target recognition task for two audiovisual datasets.
With textual records from two acoustic datasets, the proposed MVR method achieves good performance on the target localization task as measured by two regression metrics.
Case studies conducted on the ShipsEar dataset show that our MVR approach uses multi-view clustering and community assignment algorithms that outperform existing methods when constructing target relationships.
These factors make our predicted relationship graph more precise and closer to the original dataset than other methods.

In the future, we plan to extend our approach to extract underwater knowledge graphs with interpretability.
As a prerequisite, we need to investigate textual description models to represent entity properties with the aim of enabling interaction between machines and humans during knowledge inference.


\bibliography{anthology}
\bibliographystyle{IEEEtran}

\appendix

\section{Appendix}
\label{sec:appendix}

\subsection{Symbol Explanation}
The symbols we defined in this paper are listed in Tab. \ref{table5}:

\renewcommand\arraystretch{1.2}
\begin{table}[htbp]
\centering
\caption{Summary of the Notations}
\label{table5}
\resizebox{0.98\columnwidth}{!}{
\begin{tabular}{p{0.1\textwidth}<{\centering}p{0.4\textwidth}}
\hline
\multicolumn{1}{c}{\textbf{Symbols}}&\textbf{Meaning}\\
\hline
$x(t)$&A audio signal with timestamp $t$.\\
$X(k)$&The FFT of signal $x(t)$, $k$ is the frequency coordinate.\\
$g(\cdot)$&The window function of FFT.\\
$S_{m}$& The mel spectrum of the $m$th window.\\
$\eta$& The frequency response of the mel filter.\\
$C_{n}$& The MFCCs spectrum, $n$ is the number of mel filter.\\
$w$, $b$& The weight and bias of the MLP layer.\\
$z$& The embeddings of the MLP layer.\\
$\varphi$& A modality in the set $\phi$.\\
$\gamma$, $\beta$& Scaling and translation parameter counted to each modal.\\
$\mu$, $\sigma$& The mean and variance of $x_{\varphi}$.\\
$Q$, $K$, $V$&Weight matrices for query, key, and value vectors.\\
$D$&The dimension of the transfer matrix in the Transformer baseline.\\
$X_{pos}$, $X_{neg}$&The positive and negative training instance.\\
$\lambda$& A hyperparameter for balancing the size of the loss term.\\
$\digamma$&The qualitative representation to calculate the correlation.\\
$\tau$& A non-negative spread coefficient.\\
$\theta$& A uniform sample drawn from distribution $U(0,1)$, selected by $argmax(p_{\phi})$ in the forward pass\\
$x_{v}$& The multi-view representations with $v$-th view.\\
$M$, $S$& The dimension number, and the sample number in $x_{v}$.\\
$W^{v}$, $H^{v}$& Two low-dimensional decomposition matrix.\\
$H^{*}$& Unified representation in low-dimensional space.\\
$e$, $r$& The vertexes and relations in the graph model.\\
$\psi$&The qualitative representation to calculate the distance.\\
$C$& The pre-defined $C$ communities are in the graph model.\\
$\xi$, $\epsilon$&The weight and bias of multiple regression function.\\
\hline
\end{tabular}}
\end{table}

\subsection{Details of recognition and localization tasks}
To test the effectiveness of the proposed model, we combined the acoustic and visual datasets.
We merge the labels of the two audiovisual data and then retrain the model.
The recognition task is divided into acoustic and visual classification, which is easy to train and validate separately for unimodal models.
For generic models, we train the data for both modalities and use the unimodal data for detection separately.

For the localization task, we also performed a prediction of the attributes recorded when acoustic samples were collected, i.e., the underwater target localization tasks.
The details of the categories and attribute records are described in Tab. \ref{table7}:

\renewcommand\arraystretch{1.2}
\begin{table}[htbp]
\centering
\caption{Details of categories and textual records tasks in the combined audiovisual datasets.}
\label{table7}
\resizebox{0.98\columnwidth}{!}{
\begin{tabular}{ccc}
\hline
\multicolumn{1}{c}{\textbf{Dataset}}&\textbf{Specification}&\textbf{Categories}\\
\hline
\makecell[l]{DeepShip-\\SeaShip}&4-class&\makecell[l]{[Cargo, Container ship, Passenger ship, Tugboat]}\\
\hline
\makecell[l]{ShipsEar-\\SeaShip}&6-class&\makecell[l]{[Dredger, Container ship, Fish boat, Passenger ship,\\ Pilopship, Tugboat]}\\
\hline
\hline
\multicolumn{1}{c}{\textbf{Dataset}}&\textbf{Records}&\textbf{Description}\\
\hline
\makecell[l]{DeepShip-\\SeaShip}&Distance&\makecell[l]{The distance between the ship and the sensor is\\provided three times according to three different\\times, denoted as [xx-xx-xx].}\\
\hline
\multirow{8}{*}{\makecell[c]{ShipsEar-\\SeaShip}}&Hydrophone&\makecell[l]{3 device IDs.}\\
~&Gain coefficient&\makecell[l]{Recording with a hydrophone with gain coefficient.}\\
~&Submarine distance&\makecell[l]{Heights measured from the sea bottom, $\in\left[0-7.5\right]$.}\\
~&Channel Depth&\makecell[l]{Sea depth (m) at the hydrophone position, $\left[0-15\right]$.}\\
~&Wind speed&\makecell[l]{Wind speed (km/h) measured in-situ, $\left[0-18\right]$.}\\
~&Distance&\makecell[l]{Approximate distance between the vessel recorded\\and the vertical of the hydrophone.}\\
\hline
\end{tabular}}
\end{table}

\subsection{Details of model training}
We pre-trained two models, a WideMLP for text embedding and a Transformer for audiovisual feature fusion.

\textbf{WideMLP:}
The WideMLP has two ReLU-activated hidden layers with 1024 rectified linear units. 
While this may be over-parametric for nonlinear regression, it also leads to better generalization.
A dropout strategy is applied in each hidden layer, especially after the initial embedding layer.
In BoW-based WideMLP, dropout and ReLU are not used to initialize the pre-training, but only for the subsequent layers.
In optimization, we use Adam-optimized cross-entropy with a learning rate of $10^{-3}$ and a dropout rate of 0.5.

\textbf{Transformers:}
All Transformer-based models were performed on the pre-trained baseline (number of layers $L$=6, hidden dimension $D$=256, note header $h$=8).
The batch size and the epoch number are set to 8 and 128, respectively.
We adopt the strategy of early stopping and declined learning rate, i.e., the initial learning rate is $10^{-3}$ and will decrease to 60\% of the former every 10 epochs, and stop the training process if the loss reduction is less than $5*10^{-5}$ in 10 consecutive epochs.
All baselines were trained by a computer with a Core i7-8700 CPU and two Nvidia GeForce RTX 2080-Ti GPUs.



\subsection{Metrics of graphical analysis}
The graphical models are measured by computing the relationships of the nodes:

\subsubsection{Average degree}
The total number of degrees is divided by the node number in the graph. 

\subsubsection{Diameter}
The diameter, or path length, is the average distance between any two nodes.

\subsubsection{Clustering coefficient}
The clustering coefficient is the ratio of the number of edges between contact nodes divided by the theoretically maximum number of these edges.

\subsubsection{Largest Connected components}
The maximum set of connected paths traversed by breadth- or depth-first traversal.

\subsubsection{Closeness Centrality}
The closeness Centrality is used to calculate the sum of distances from a node to all other nodes.

\subsubsection{Betweenness Centrality}
The percentage of shortest paths of other node pairs in the graph that pass through the node. 

\subsubsection{Transitivity}
A fraction of all possible triangles in the diagram, these possible triangles are identified by the number of "triangles" (two edges with a shared vertex).

\subsubsection{Effective Size}
The effective size is the non-redundant part of the node relationship.




\end{document}